\documentstyle[aps,prb,preprint]{revtex}
\begin{document}
\title
{Anomalous Charge Dynamics in the Superconducting State of Underdoped
Cuprates}

\author{A.J. Millis}
\address{Department of Physics and Astronomy,
The Johns Hopkins University,
3400 North Charles Street,
Baltimore, MD  21218}
\author{S. M. Girvin}
\address{Department of Physics,
Indiana University,
Swain Hall W 117,
Bloomington, IN  47405}
\author{L. B. Ioffe}
\address{Department of Physics,
Rutgers University,
Piscataway, NJ  08544}
\author{A. I. Larkin}
\address{Department of Physics,
University of Minnesota,
Minneapolis, MN 55455}
\maketitle
\begin{abstract}

We present fermi liquid expressions for the low temperature behavior of the
superfluid stiffness, explain why they differ from those suggested recently
by Lee and Wen, and discuss their applicability to data on high-$T_c$
superconductors.  We find that a consistent description requires a strong,
doping dependent anisotropy, which affects states near the zone corners
much more strongly than those near the zone diagonals.
\end{abstract}
\newpage

     The work reported here was motivated by a recent paper of Lee and Wen
on "Unusual Superconducting State of Underdoped Cuprates" \cite{Lee97}.
Their point of view, which we share, is 
that because the superconducting state of the high-$T_c$
materials is apparently conventional and in particular supports
well-defined quasiparticle excitations, it is appropriately  described 
in terms of fermi liquid theory. Their main conclusion, that
the temperature dependence of the superfluid stiffness is anomalous
and implies unusual properties, is important and interesting. Their
specific results must be interpreted with caution because
their formulation of fermi liquid theory in the superconducting state (Eqs
1-4 of their paper) disagrees in an important way with the standard
formulation \cite{Larkin64}.  In this note we present the usual
fermi liquid theory of the charge response of the superconducting state,
explain why it differs from Eqs 1-4 of Lee and Wen, and apply it to
penetration-depth data and discuss the underlying physics.

     We begin, however, with data.  We write the penetration
depth $\lambda$, in
terms of a superfluid stiffness per $CuO_2$ plane $\rho_s$ via $\rho_s =
\hbar^2 c^2b/16\pi e^2 \lambda ^ 2$.  Here b is the mean interplane
spacing.  
This definition yields the coefficient of the $\frac{1}{2}(\nabla \theta) ^2$
term in the Ginzburg-Landau theory but differs by a factor of 4 from
the quantity defined in \cite{Lee97}.  The most
extensively studied family of materials is the `YBCO' family; to avoid
complications due to the $CuO_2$ chains we consider the penetration depth
for current flow in the basal plane but perpendicular to the chains.  In
YBCO the length b  $\approx 5.5 \AA$.
The measured \cite{Bonn95} 
penetration depths are $\lambda_{aa} =1600$ for $YBa_2Cu_3O_7$ and
$\lambda_{aa} = 2000$ for $YBa_2Cu_3O_{6.6}$ 
implying $\rho_{saa} = 11meV$ 
($O_7$) and $\rho_{saa} = 7 meV$ ($0_{6.6}$).  In a clean weak-coupling
superconductor $\rho_s$ may be determined from the conduction band plasma
frequency $\omega_p$ via $\rho_s = (\hbar \omega_p) ^ 2b/(16\pi \
e^2)$; the value 
calculated using band theory is \cite{Orenstein90}
$(\hbar \omega_p ) ^2= 9 eV^2$ 
implying $\rho_{saa-band}= 70 meV$.
The renormalization of $\rho_s$ from the band theory prediction is
large (factor of 6-10) and doping dependent \cite{TKT}. 

The temperature dependence of
$\rho_s$ for YBCO is given in \cite{Bonn95}.  For both $O_7$ and $O_{6.6}$
$\rho_s$ is well fit by $\rho_s(T) = \rho_s (T=O) (1-0.5T/T_c)$.  
Converting the data to absolute units we
find $\lim_{T\rightarrow 0} d\rho_s/dT = 0.06 meV/K$ for both materials.
The theoretical result \cite{Lee97} is
$d\rho_s/dT = (ln2 / 2\sqrt{2}) (v_F/a) /d\Delta/d\theta$  where
$d\Delta/d\theta$ is the angular derivative of the gap function at the gap
node and $a=3.8 \AA$ is the lattice constant.
Photoemission results \cite{Harris96} on BSCCO show that if the gap is
defined as the displacement of the midpoint of the leading edge of the
photoemission spectrum from the fermi level, then $d\Delta/d\theta\approx
40 meV \approx 450K$; using the band theory $v_F \approx 4 eV-\AA$
 yields $d\rho_s/dT =
0.5  meV/K$.

     We now interpret these data in terms of fermi liquid theory.  Fermi
liquid theory is specified by the quasiparticle velocity $\vec{v_F}
(\theta)$ and the Landau interaction function $f (\theta, \theta')$; these
are functions defined on the fermi surface which for the electronically
two-dimensional high-$T_c$ materials is a simple closed curve $p_F(\theta)$
with $\theta$ the usual polar angle (we neglect the chain bands and any
bonding-antibonding splitting arising from the bilayer structure).
 
A basic assumption of the
fermi liquid analysis is that the low-T dependence is due to
thermally excited quasiparticles.  Several authors have 
claimed \cite{Emery95,Roddick95}
that  phase fluctuations of the superconducting order parameter
can also lead to such a T-dependence.  However,
these models require 
a large density of normal electrons at $T=0$ to screen the
Coulomb interaction.  In high-$T_c$ materials there are
no such electrons at $T=0$;
the phase fluctuations will therefore be gapped, giving an exponentially
small contribution.  Even if screening
is included, we believe the models of refs \cite{Emery95,Roddick95}
yield a $T^2$ temperature dependence. Claims to the contrary 
seem to have arisen from Eq
7 of ref \cite{Chakravarty86}
in which a term involving
the Matsubara sum $\sum_{n=-\infty}^{n=\infty}$ was written as $2*\sum_{n=0}^{\infty}$
so that the $n=0$ term was counted twice.  Evaluation of
$\sum_{n=-\infty}^{n=\infty}$ yields a $T^2$ temperature dependence

     The fermi velocity may be determined from the dispersion of
quasiparticle peaks observed in photoemission.  The interpretation of
photoemission data on high-$T_c$ materials is still controversial, but
there is general agreement that for momenta near the zone diagonal,
well-defined peaks exist from which a dispersion can be extracted.  The
most detailed studies have focused on the BSCCO family of materials; the
data for YBCO seem quite similar \cite{Campuzano90}.  
Fig. 3 of ref \cite{Marshall96} 
shows the dispersion along
the zone diagonal for overdoped, optimally doped, and underdoped 
BSCCO samples.
The data imply a {\it doping-independent}
quasiparticle velocity $v = 1.3eV -
\AA$; this is reduced from the
band-structure value ($v_F \approx 4 eV)$ by about a factor of 3,
presumably because of electron-electron interaction effects.  
It is conventionally assumed (and has been demonstrated in some
models \cite{Ioffe89,Grilli90}) 
that the many-body renormalizations which change the
zone-diagonal velocity by a factor of three are relatively isotropic around
the fermi surface, so 
we might expect that the velocity everywhere to be changed by about the same
factor.  This, combined with the observation that the charge response
should be dominated by the zone diagonals where the velocity is largest,
suggests that if the Landau parameter contribution were negligible $\rho_s$
would be renormalized by about a {\it doping-independent} factor of three
from the band theory prediction.  The smaller value 
and especially the doping dependence $\rho_s$  therefore suggests that
the Landau interaction function plays an important role as
predicted theoretically \cite{Ioffe89,Grilli90}.  We now show that
while the Landau function may be important, the
conventional picture makes predictions for $d\rho_s/dT$ which are ${\it
inconsistent}$ with the data.

     The fermi liquid theory of the superconducting state was developed in
\cite{Larkin64}.  The result may be written:

\begin{equation}
\rho_{sab} = \frac{\hbar ^ 2 b}{2\pi^2} \int{d\theta_1 d\theta_2
N(\theta_1) N(\theta_2) v_a (\theta_1) L(\theta_1) T_b (\theta_1,
\theta_2)}
\label{rho}
\end{equation}

Here $N(\theta) = 1/v(\theta)$ is the quasiparticle density of states at
position $\theta$ on the fermi surface, $v_a$ is a component of the
quasiparticle velocity and $L(\theta) = \pi T \sum_n \Delta(\theta
\omega_n)^2 / [\omega_n^2 + \Delta(\theta, \omega_n)^2] ^ {3/2}$ is the
usual angle-dependent Yoshida function.  The vertex $\vec{T}$ obeys the
equation

\begin{equation}
\vec{T} = \vec{v} + \langle f L \vec{T}\rangle
\label{T}
\end{equation}

Here $f$ is the Landau function defined above and products
inside angle brackets are shorthand
for integrals over the fermi line, weighted by density of states as in Eq.
\ref{rho}.  Eq. \ref{T} may be solved formally and the solution used in
Eq. \ref{rho}.  Noting that at $T=0$ $L=1$ and that the leading T-dependence
is in L one finds:

\begin{equation}
\rho_{saa} (T=0) = <v_a(1-f)^{-1} v_a>
\label{rho0}
\end{equation}

\begin{equation}
\lim_{T\rightarrow 0} d\rho_{saa}/dT = <v_a(1-f) ^{-1} dL/dT(1-f) ^{-1}v_a>
\label{drhodT}
\end{equation}

Eq. \ref{drhodT} differs from that written in ref \cite{Lee97} by the two
factors of $(1-f) ^{-1}$.  The authors of \cite{Lee97} 
obtained their result by analogy to
Leggett's theory of the spin susceptibility of $^3He$; the results of
\cite{Larkin64} show the analogy is not appropriate.  The physics is this:
Landau parameters enter expressions for physical quantities because a
perturbing field H excites quasiparticle-quasihole pairs at all points on
the fermi line; i.e. it modifies the underlying fermi distribution and
because of the electron-electron interaction this
changes the quasiparticle energy dispersion to order H, and thus affecting
the number of excitations.  This effect must be treated self-consistently.
Now in $^3He$ the condensate does not respond at linear order to a magnetic
field (at least in the low-T "BW" phase), and because of the
superconducting gap the number of quasiparticles at low T is very small;
thus at low T changes to the quasiparticle dispersion may be neglected and
Landau parameters drop out.  By contrast a superconducting condensate does
respond to a magnetic field (by producing a supercurrent); this supercurrent
does correspond to a displacement of the underlying fermi distribution
which does affect the superconducting quasiparticles, causing Landau
parameter effects in the temperature dependence.

     In fermi liquid theory one defines the Landau
parameters $A_{1s}$, $F_{1s}$ via $<v_a(1-f)^{-1} v_a> = <v_a^2>/(1-A_{1s}/2) =
<v_a^2> (1+F_{1s}/2)$.  If the renormalization of the velocity is
isotropic over the fermi surface we may
evaluate $<v_a^2>$ by dividing the band theory $\rho_s$ by three
(recall the density of states factor in the definition of $< >$) to obtain $<v_a^2>=23 meV$.  By comparing this to the data we
conclude that
$(1+F_{1s}/2) \approx 1/2$ ($O_7$) and $(1+F_{1s}/2) \approx 1/4$ ($O_{6.6}$).
Now in a Galilean-invariant fermi liquid 
$v_F=p_F/m^*$ and $<v_a^2> \approx n/m^*$.  In a
two dimensional Galilean-invariant
system the Landau parameter obeys $1+F_{1s}/2=m^*/m$ so the
Landau parameter cancels the mass renormalization.  In non-Galilean-invariant
systems such as the high-$T_c$ superconductors there is no a-priori
relation between $F_{1s}$ and $m^*/m$.  As we see,
in the actual materials the observed $\rho_s$ is smaller
than one would expect from the observed $v$; thus $F_{1s}$ acts in the same
direction as $m^*$, instead of compensating it.  

     In many fermi liquids the angular dependence of the
interaction is not too strong, so one expands the operator f in appropriate
harmonics and retains only the one most closely corresponding to $\vec{v}
\vec{v}$; this amounts to replacing the operator $(1-f) ^ {-1}$ by the
scalar $1+F_{1s}/2$ in both eqs. \ref{rho0} and \ref{drhodT}.  
We refer to this as the ``conventional fermi liquid'' approximation.
A particular realization of such a fermi liquid
is found in the slave boson calculations of \cite{Ioffe89,Grilli90}.
To compute $d\rho_s/dT$ in this
conventional fermi liquid approach we
take the "band theory" value of $0.5 meV/K$ obtained earlier, divide
by a factor of three for the velocity renormalization and a further factor
of 4 for $YBa_2Cu_3O_7$ and 16 for $YBa_2Cu_3O_{6.63}$, yielding $d\rho_s/dT
= 0.04 meV/K$ for $O_7$ and $0.01 meV/K$ for $O_{6.6}$.  Although the
estimate of $d\rho_s/dT$ for $O_7$ is within a factor of two of the measured
value of $0.06meV/K$,  the
estimate for $O_{6.6}$ is badly off.  Similarly, ref \cite{Ioffe89} found $\rho_s(T)=<vL(T)v>/(1+c<vL(T)v>$
with $c \approx -A_{1s} \sim 1/x$, so $d\rho_s/dT \sim \rho_s(T=0)<vdL/dTv>/c
\sim 1/x^2$.   Thus both the  conventional fermi liquid  approach 
and explicit calculations predict
a strong doping dependence where none exists experimentally.
In ref \cite{Lee97} the disagreement of the result of ref \cite{Ioffe89} 
with the observed $d\rho_s/dT$ was argued to indicate the inadequacy
of the $U(1)$ gauge theory formalism used in
\cite{Ioffe89}.  The present analysis makes
it clear that the difficulty is more general.

     The doubtful point of the discussion given above is the assumption
of weak angular dependence around the fermi surface.  This enters the
argument in two places: first, that the factor-of-three
velocity renormalization observed for quasiparticles along the zone
diagonal applies all over the fermi surface, and second that the Landau
interaction operator $(1-f)^{-1}$ could be replaced by the Landau parameter
$1+F_{1s}/2$. These two issues are clearly related: 
an interaction giving a relatively
larger velocity renormalization at the zone corners than at the zone
diagonals will lead to a Landau function with a stronger angular
dependence. 

If the velocity renormalization increases rapidly as one moves
away from the zone diagonal then the estimate of the renormalization of
$\rho_s$ relative to band theory  increases and the value of the Landau
parameter decreases.  There is evidence of anomalous flatness of bands
and shortness of lifetimes in the vicinity of the zone corners
\cite{Dessau93}.  In order to explain the data the
size of the "non-flat" regions must be small and x-dependent.
Evidence on this point is not conclusive.

Consider now the possibility of
strong anisotropy in the Landau function.
We see from Eq. \ref{drhodT} that because the
quasiparticle excitations are concentrated in the nodes, at low T the
temperature-dependent quantity $dL/dT$ is non-zero only near the zone
diagonal. If the Landau function $f(\theta_1,\theta_2)$
were anomalously small for either $\theta_1$ or $\theta_2$ near this point
then the effect of the Landau function on $d\rho_s/dT$ would
be greatly reduced in agreement with data.
The range of angles over which this occurs must be small and
f must be
x-dependent.

To summarize:  the temperature dependence of the superfluid
stiffness strongly suggests that some process acting in the high-$T_c$
superconductors strongly suppresses the contribution of the particles
away from the zone diagonal to the current.

{\it Acknowledgements} We thank E. Abrahams, I. Aleiner and B. G. Kotliar
for helpful conversations.  L. B. I., A. I. L. and A. J. M. thank
NEC research for hospitality.  S. M. G. and A. J. M. thank the
Aspen Center for Physics.  S. M. G. was supported by D.O.E.
grant DE-FG02-90ER45427 and A. J. M. by NSF DMR-9707701.


\begin{thebibliography} {99}

\bibitem{Lee97} P. A. Lee and X. G. Wen, Phys. Rev. Lett. {\bf 78} 4111 (1997).

\bibitem{Larkin64} A. I. Larkin, J. Exptl. Theoret. Phys. (U.S.S.R.) {\bf
46} 2188 (1964) Sov. Phys. J.E.T.P. {\bf 46} 1478 (1965) and
A. J. Leggett, Phys. Rev. {\bf 140} A1869 (1965); see also F.
Gross et. al., Z. Phys. {\bf B64} p. 175 (1986) and 
D. Xu, S. K. Yip and J. A. Sauls, Phys. Rev. {\bf B51} 16233
(1995).

\bibitem{Bonn95} D. A. Bonn et. al., J. Phys. Chem. Sol. {\bf 56} 1941 (1995).

\bibitem{Orenstein90} J. Orenstein et. al., Phys. Rev. {\bf 42} 6342 (1990).
This reference discusses band structure calculations and their comparison
to optical and penetration depth data.

\bibitem{TKT} It should be noted
however that although the $\rho_{saa}$ values we have quoted  
seem quite small, the phase stiffness controlling a possible
Kosterlitz-Thouless temperature is much larger.  First,
one must use $(\rho_{saa}\rho_{sbb})^{1/2}$ (the b direction
is parallel to the chains); second the YBCO materials have a bilayer structure
in which pairs of planes are strongly coupled, implying a further
factor of two increase.  Combining these factors we find that the
Kosterlitz-Thouless transition temperature implied by the $T=0$ $\rho_s$ is
$T_{KT}\equiv \pi \rho_s(T=0)/2$ $=600K$ ($O_7$) and $300K$ $O_{6.6}$,
much greater than the observed $T_c$ of these materials.


\bibitem{Emery95} V. J. Emery and S. Kivelson, Nature, {\bf 374}, p. 434-7
(1995).

\bibitem{Roddick95}  E. Roddick and D. Stroud,
Phys. Rev. Lett., {\bf 74},  p. 1430-3 (1995).

\bibitem{Chakravarty86} S. Chakravarty et. al., Phys. Rev. Lett. {\bf 56} 2303 (1986).

\bibitem{Campuzano90} J. Campuzano et. al., Phys. Rev. Lett. {\bf 64}
2308 (1990).

\bibitem{Marshall96}  D.S. Marshall, D. S. Dessau, A. G. Loeser, C-H Park,
A. Y. Matsuura, J. N. Eckstein, I. Bozovic, P. Fournier, 
A. Kapitulnik, W. E. Spicer
and Z-X Shen, Phys. Rev. Lett. {\bf 76} 4841 (1996).

\bibitem{Ioffe89} L. B. Ioffe and A. I. Larkin, Phys. Rev. {\bf B39} 8988
 (1989).

\bibitem{Grilli90} M. Grilli and B. G. Kotliar, Phys. Rev. Lett. {\bf 64}
1170 (1990) and B. G. Kotliar, p. 197 in {\it Les Houches Session LVI 1991},
Elsevier (1995).

\bibitem{Harris96} J. M. Harris, Z.-X. Shen, P. J. White, D. S. Marshall,
M. C. Schabel, J. N. Eckstein and I. Bozovic, Phys. Rev. {\bf B54} 15665
(1996).

\bibitem{Dessau93} D. S. Dessau et. al, Phys. Rev. Lett. {\bf 71} 2781 (1993).

\end{thebibliography}
\end{document}